\documentclass[prb,twocolumn] {revtex4-1} 

\usepackage{amsmath}  
\usepackage{amsfonts} 
\usepackage{graphicx} 
\usepackage{longtable}  

\newcommand{\BLV}{\ensuremath{(BL)_\mathrm{V}}}
\newcommand{\BLF}{\ensuremath{(BL)_\mathrm{F}}}

\begin{document}

\title{A LEGO Watt Balance: \\An apparatus to determine a mass based on the new SI}

\author{L. S. Chao}
\affiliation{Physical Measurement Laboratory, National Institute of Standards and Technology, Gaithersburg, MD 20899}

\author{S. Schlamminger}
\affiliation{Physical Measurement Laboratory, National Institute of Standards and Technology, Gaithersburg, MD 20899}

\author{D. B. Newell}
\affiliation{Physical Measurement Laboratory, National Institute of Standards and Technology, Gaithersburg, MD 20899}

\author{J. R. Pratt}
\affiliation{Physical Measurement Laboratory, National Institute of Standards and Technology, Gaithersburg, MD 20899}

\author{F. Seifert}
\affiliation{Joint Quantum Institute, University of Maryland, College Park, MD 20742}

\author{X. Zhang}
\affiliation{Joint Quantum Institute, University of Maryland, College Park, MD 20742}

\author{G. Sineriz}
\affiliation{Joint Quantum Institute, University of Maryland, College Park, MD 20742}

\author{M. Liu}
\affiliation{Joint Quantum Institute, University of Maryland, College Park, MD 20742}

\author{D. Haddad}
\affiliation{Joint Quantum Institute, University of Maryland, College Park, MD 20742}

\date{\today}

\begin{abstract}
A global effort to redefine our International System of Units (SI) is underway and the change to the new system is expected to occur in 2018. Within the newly redefined SI, the present base units will still exist but be derived from fixed numerical values of seven reference constants. In particular, the unit of mass, the kilogram, will be realized through a fixed value of the Planck constant $h$.  A so-called watt balance, for example, can then be used to realize the kilogram unit of mass within a few parts in $10^8$. Such a balance has been designed and constructed at the National Institute of Standards and Technology. For educational outreach and to demonstrate the principle, we  have constructed a LEGO tabletop watt balance capable of measuring a gram-size mass to 1\% relative uncertainty.  This article presents the design, construction, and performance of the LEGO watt balance and its ability to determine~$h$. 
\end{abstract}

\maketitle

\section{Introduction} 

The quest for a redefined International System of Units (SI) has been a formidable global undertaking. If the effort concludes as expected, sometime in 2018 the seven base units (meter, kilogram, second, ampere, kelvin, mole, candela) that have formed the foundation of our unit system for over half a century will be redefined via seven reference constants. In terms of mass metrology, the present standard, forged in 1879 and named the International Prototype Kilogram (IPK), is the only mass on Earth defined with zero uncertainty. In the redefined system, the base unit kilogram will be redefined via a fixed value of the Planck constant $h$, finally severing its ties to the IPK.  Different experimental approaches can be used to realize~\cite{realization} mass from the fixed value of $h$. At the National Institute of Standards and Technology (NIST), we have chosen to pursue the watt balance to realize the kilogram in the US after the redefinition.~\cite{Seifert14}

The watt balance, first conceived by Dr.~Bryan Kibble in 1975, is a mass metrology apparatus that balances the weight of an object against an electromagnetic force generated by a current-carrying coil immersed in a magnetic field. By design, the watt balance toggles between two measurement modes and indirectly compares
electrical power and mechanical power, measured in units of watts---hence the term ``watt balance.''\cite{Kibble75} It is essentially a force transducer that can be calibrated solely in terms of electrical, optical, and frequency measurements. 
A few watt balances around the world have demonstrated the capability of measuring 1~kg masses with a relative uncertainty of a few parts in $10^8$.~\cite{Stock13}

Here, under the inspiration of Terry Quinn,~\cite{Quinn13} we describe the construction of a tabletop LEGO~\cite{Note1} 
watt balance capable of measuring gram-level masses with a much more modest relative standard uncertainty of 1\%. For the instrument described here, the cost of parts totaled about \$650, but a similar device can be built for significantly less. The largest portion of the cost is in the data acquisition system used to transfer the data to a computer. A recommended parts list is provided in Appendix~A. We encourage readers to use this manuscript as general guidance for constructing such a device and by no means as a definitive prescription. There are many ways to build a watt balance, and we consider here a concept to highlight general considerations that are most important for success.

\section{Basic Watt Balance Theory}

Although we understand that the reader is eager to hear about the LEGO watt balance, we will first explain the physics underpinning the professional watt balance. Several national metrology institutes worldwide have constructed watt balances and are presently pursuing ultra high-precision mass measurements. These watt balances can measure masses ranging from 500\;g to 1\;kg and obtain relative standard uncertainties as small as a few parts in $10^8$, or about a million times smaller than that of the LEGO watt balance. 

\begin{figure}[h!]
 \centering
 \begin{minipage}{.5\textwidth}
   \centering
   \includegraphics[width=8.5cm]{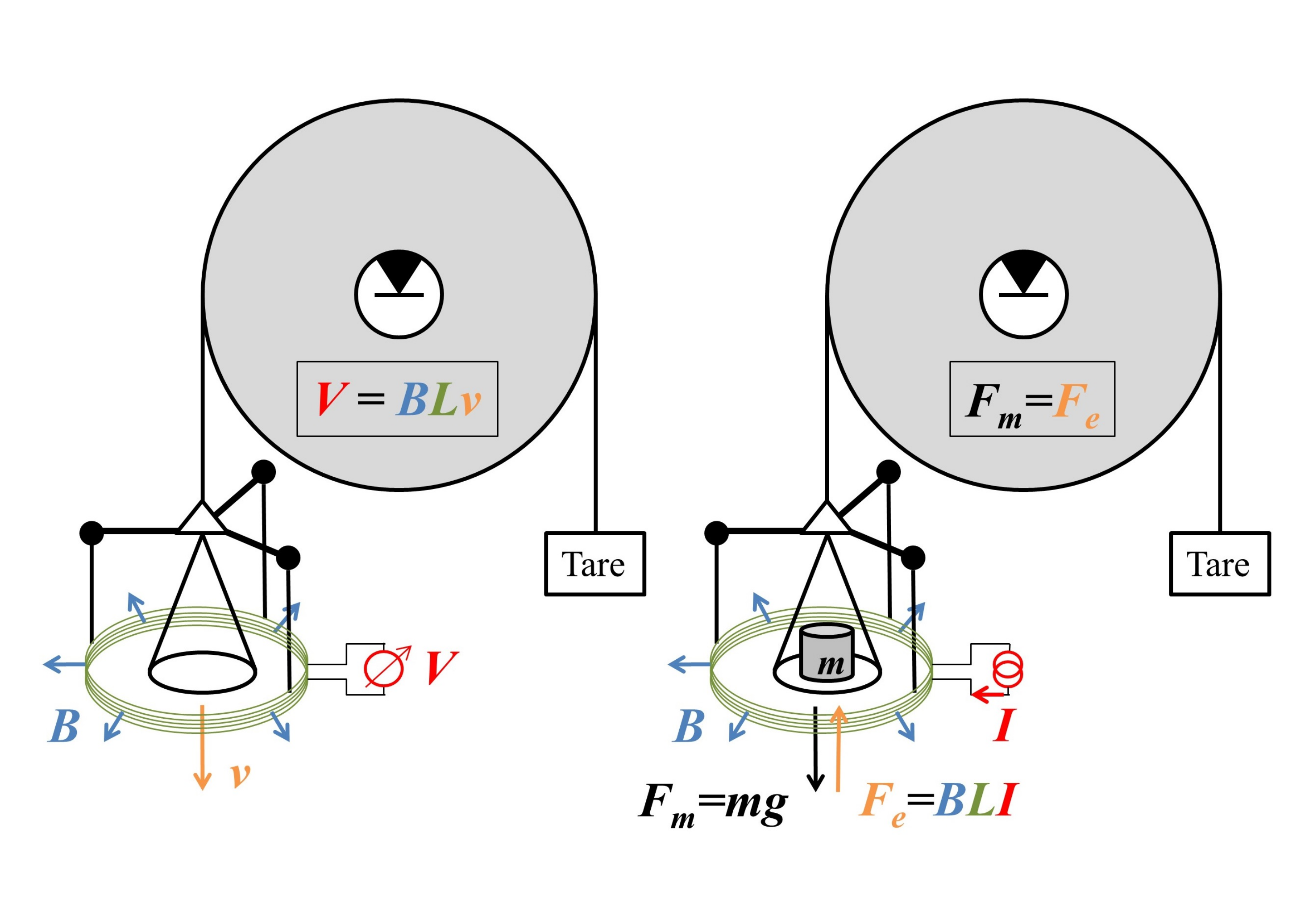}
 \end{minipage}%
 \caption{Left: velocity mode. The coil moves vertically in a radial magnetic field and a voltage $V$ is induced. Right: force mode. The upward electromagnetic force generated by the coil opposes the gravitational force exerted by $m$.}
 \label{fig:theory}
 \end{figure}

Even though a watt balance might appear functionally similar to an equal-arm balance, an equal-arm balance is passive, relying on comparing an unknown mass to a calibrated one, while a watt balance is active, relying on compensating the unknown weight with a known force. In this case, the weight of an object is compensated by a precisely adjusted electromagnetic force. The experiment involves two modes of operation, illustrated in Fig.~\ref{fig:theory}:  velocity mode and force mode. Velocity mode is based on the principle of Lorentz forces. A coil (wire length $L$) is moved at a vertical speed $v$ through a magnetic field (flux density $B$) so that a voltage $V$ is induced. The induced voltage is related to the velocity through the flux integral $BL$: 
\begin{equation}
V=BLv.
\end{equation}
Similarly, force mode is also based on Lorentz forces. The gravitational force on a mass $m$ is counteracted by an upward electromagnetic force $F$ generated by the now-current-carrying coil in a magnetic field:
\begin{equation}
F= BLI = mg,
\end{equation}
where $g$ is the local gravitational acceleration and $I$ is the current in the coil.

In principle, mass could be realized solely by operating in force mode---if $B$ and $L$ could be measured accurately. Because both of these variables are difficult to measure precisely, velocity mode is necessary as a calibration technique. By combining Eqs.\ (1) and~(2), canceling out the $BL$ factor common to both equations, and rearranging the variables, expressions for electrical and mechanical power are equated and a solution for mass is obtained:
\begin{equation}
VI=mgv \Longrightarrow m = \frac{VI}{gv}. \label{eq:watt}
\end{equation}

The equation above relates mechanical power to electrical power and provides a means to relate mass to electrical quantities. The relationship equates ``virtual'' power, in the sense that the factors of each product, $V$ and $I$ or $mg$ and $v$, are not measured simultaneously, but separately in the two modes. The ``power'' only exists virtually, i.e., as a mathematical product. The practical significance of a ``virtual'' comparison is that the result is independent of several friction terms, such as the mechanical friction during velocity mode or the electrical resistance of the coil wire.
 
In order to make the connection from mass to the Planck constant through the electrical quantities, it is necessary to understand two quantum physical effects that have revolutionized electrical metrology since the second half of the last century: the Josephson effect and the quantum Hall effect. These two phenomena are what permit the measurement of electrical quantities in terms of the Planck constant to the precision required for the watt balance and redefinition. On a side note, another constant, the elementary charge $e$, is present in both the Josephson effect and the quantum Hall effect. However, in the final watt balance equation, the elementary charge drops out.

The Josephson effect can be observed in a Josephson voltage standard, which consists of two superconducting materials separated by a thin non-superconducting barrier. At superconducting temperatures, and while irradiating the junction with an electrical field at a microwave frequency $f$, a bias current is forced through this junction and a voltage of
\begin{equation}
V=\frac{h}{2e} f \equiv K_\mathrm{J}^{-1} f
\end{equation}
will develop across the junction. The quotient $K_\mathrm{J} = 2e/h$  is named the Josephson constant in honor of Brian Josephson, who predicted this effect in 1960.~\cite{Clarke70} One junction delivers only a small voltage, typically 37\,$\mu$V, so, in order to build a practical voltage standard, tens of thousands of these junctions are connected in series on a single chip.  At NIST,~\cite{Tang12} a chip the size of an index card with approximately 250,000 junctions is immersed in liquid helium and can produce any voltage up to 10\,V with an uncertainty of 1\,nV.  In principle, the Josephson voltage standard is a digitally adjustable battery---with a $\approx\,$\$100,000 price tag.

The quantum Hall effect is a special case of the Hall effect. The Hall effect occurs when a current-carrying conductor is immersed in a magnetic field and a Hall Voltage $V_H$ occurs perpendicular to the magnetic flux and the current. While in the classical Hall effect the conductor immersed is a three-dimensional object, in the quantum Hall effect, the electrical conduction is confined to two  dimensions. In such a system and at sufficiently high magnetic field, the ratio between the Hall voltage and current, or Hall resistance $R_H$, becomes quantized to 
\begin{equation}
R_H=\frac{V_H}{I}=\frac{1}{i}\frac{h}{e^2}\equiv \frac{1}{i} R_\mathrm{K},
\end{equation}
where $i$ is an integer.  The quotient $R_\mathrm{K} = h/e^2$ is named the von Klitzing constant to honor Klaus von Klitzing, who discovered this effect first in 1980 (see Ref.~\onlinecite{Eisenstein92}). At NIST, the quantum Hall effect is the starting point of resistance  dissemination.~\cite{Rand01} Scaling with a cryogenic current comparator allows researchers to measure a  $100\;\Omega$ precision resistor with a relative uncertainty of a few parts in $10^9$. On the outside, a quantum Hall system looks similar to a Josephson voltage system: a bundle of cables leading into a liquid helium dewar. On the inside, a fingernail-sized chip sits in a strong magnetic field at temperatures below 1.5\,K. A skilled operator can use the device to realize the same resistance value independent of time and place. 

Together, these two quantum electrical standards enable scientists at NIST to build a watt balance with a relative measurement uncertainty that is about 1 million times smaller than that of the LEGO watt balance built at home or in the classroom.
You may be wondering why all of a sudden we need to make a resistance ($R$) measurement when we actually need a current ($I$) measurement. Because a high-precision measurement of $I$ is difficult to achieve, we simply use Ohm's Law and equate $I=V/R$. Hence, instead of measuring $P=VI$, the current $I$ is driven through a precisely calibrated resistor $R$, producing a voltage drop $V_R$, yielding $P=V V_R/R$. Both voltages are measured by comparing to a Josephson voltage standard, so their values can be expressed in terms of a frequency and the Josephson constant. The resistor is measured by comparing to a quantum Hall resistor, so its value can be expressed in terms of $R_K$. This can be written as
\begin{equation}
P = V V_R/R = C f_1 f_2 \frac{h}{2e}  \frac{h}{2e} \frac{e^2}{h}  = \frac{C f_1 f_2}{4} h.
\end{equation}
Here, $C$ is a known constant that indicates the number of junctions used and the ratio of $R$ to $ h/e^2$. Combining the above equation with Eq.~(\ref{eq:watt}) yields
\begin{equation}
h = \frac{4}{C f_1 f_2} mgv    \Longrightarrow m = \frac{C f_1 f_2}{4} \frac{h}{gv}.
\end{equation}
Before the 2018 redefinition of units, the equation on the left is used to measure $h$ from a mass traceable to the IPK. After redefinition, the equation on the right will be used to realize the definition of the kilogram from a fixed value of $h$ in joule-seconds. 

In a classroom setting, quantum electrical standards are typically unavailable. However, it is still possible to measure the Planck constant, due to the way the present unit system is structured. While the SI is used for most measurements, a different system of units has been used worldwide for almost all electrical measurements since 1990. For these so-called conventional units, the Josephson and von Klitzing constants were fixed at values adjusted to the best knowledge in 1989.~\cite{Quinn89,Taylor89} These fixed values are named ``conventional Josephson" and ``conventional von Klitzing" constants and are abbreviated $K_\textrm{J-90}$ and $R_\textrm{K-90}$, respectively. Since 1990, almost all electrical measurements are calibrated in conventional units. By comparing electrical power in conventional units to mechanical power in SI units, $h$ can be determined. 

Starting at Eq.~(\ref{eq:watt}), we see that
\begin{equation}
VI = mgv  \Longrightarrow \{VI\}_{90} \mathrm{W}_{90} = \{mgv\}_{\mathrm{SI}}  \mathrm{W}_{\mathrm{SI}},
\end{equation}
where $\{x\}_{90}$ and $\{x\}_\textrm{SI}$ denote the numerical values of the quantity $x$ in conventional and SI units, respectively. Further, $\mathrm{W}_{90}$ and $\mathrm{W}_{\mathrm{SI}}$ are the units of power (watt) in the conventional and SI systems. The equation above can be written as
\begin{equation}
\frac{  \{mgv\}_{\mathrm{SI}} }{ \{VI\}_{90} } = \frac  {  \mathrm{W}_{90} }{  \mathrm{W}_{\mathrm{SI}} } = \frac{h}{h_{90}}
 \Longrightarrow  h = h_{90} \frac{  \{mgv\}_{\mathrm{SI}} } { \{VI\}_{90} },
\end{equation}
where $h_{90}$ is the conventional Planck constant, defined as
\begin{equation}
h_{90} \equiv \frac{4}{{K_\textrm{J-90}^2}R_\textrm{K-90}}  =  6.626\,068\, 854\,\ldots\times 10^{-34}\,\mbox{J\,s}.
\end{equation}
Thus, the value of the Planck constant can be determined by multiplying the conventional Planck constant by the ratio of mechanical power in SI units to electrical power in conventional units.

To arrive at this ratio, we start by assigning different flux integrals $BL$ to each mode, i.e.,
\begin{equation}
\BLV =\frac{V}{v} \qquad \mbox{and}\qquad \BLF =\frac{mg}{I} .
\end{equation}
Using these two numbers, the ratio of $h/h_{90}$ is given by
\begin{equation}
 \frac{h}{h_{90}} = \frac{\BLF}{\BLV} = \frac{  \{mgv\}_{\mathrm{SI}} } { \{VI\}_{90} }.\label{eq:bl_comb}
\end{equation}

After redefinition, electrical power and mechanical power will be measured in the same units and the schism between units will vanish. Then, referring back to Eq.~(\ref{eq:watt}), an arbitrary mass can be determined using a watt balance simply as:
\begin{equation}
m = \frac{VI}{gv},
\label{eq:lastwatt}
\end{equation}
where all quantities are expressed in SI units.

The remaining two variables $g$ and $v$ are measured accurately by NIST scientists with an absolute gravimeter and interferometric methods, respectively. However, since this manuscript's main focus is still a proof-of-principle LEGO watt balance, ultra-high-precision metrology approaches are unnecessary. Gravity can be estimated by inputting one's geographical coordinates into the web page found in Ref.~\onlinecite{noaa}, or even measured experimentally with a simple pendulum in the laboratory.  Velocity can be determined using a simple optical method that we describe in Sec.~V.

However, do not be fooled by our toy. The LEGO watt balance is versatile and fully capable of measuring in either mode. It will be a device to measure the Planck constant before redefinition and one to realize mass after redefinition. A capable operator can perform a measurement with a relative uncertainty of 1\% with the device described below.

\section{LEGO Watt Balance Mechanics}

\begin{figure}[h!]
\includegraphics[width=8.5cm]{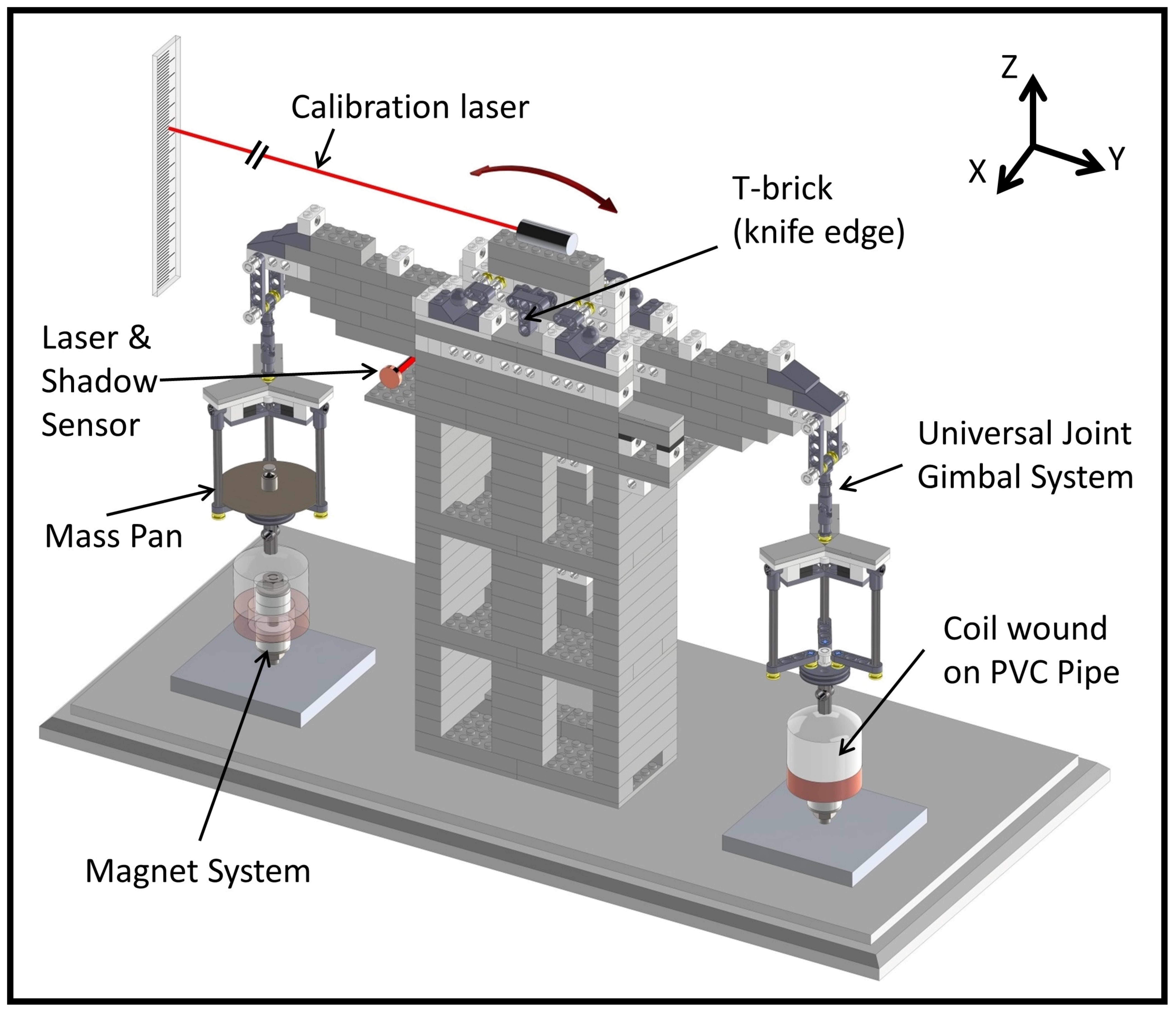}
\caption{CAD model of the LEGO watt balance. The balance pivots about the T-block at the center. Two PVC endcaps with copper windings hang from universal joints off either side of the balance beam. Coil A is on the left and Coil B is on the right. A 10 gram mass sits on the Coil A mass pan and each coil is concentric to its own magnet system. Two lasers are used to calibrate and measure the linear velocity of each coil.}
\label{fig:app}
\end{figure}

We chose a symmetric design for the LEGO watt balance that conforms to easily recognized notions of an equal-arm beam balance. We reiterate that there are many ways to construct a watt balance. One way is described below. Figure~\ref{fig:app} shows a CAD drawing of our balance. A weighing pan is suspended from each arm of the balance, which pivots about its center. Suspended below each weighing pan is a wire-wound coil immersed in a radial magnetic field.

\begin{figure}[h!]
\centering
\includegraphics[width=8.5cm]{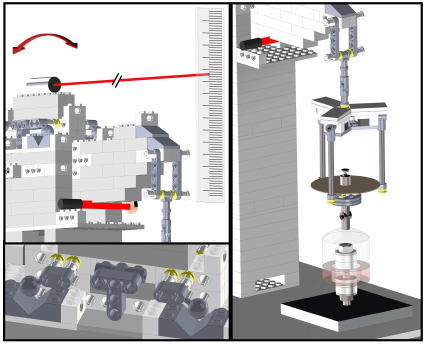}
\caption{Top left: The calibration laser projects onto a ruler a few meters away. The shadow sensor detects angular motions of the balance and outputs an oscillatory voltage signature. Right: A transparent view of the PVC coil assembly show its concentricity to the magnet assembly. A stainless steel 10~g mass sits centered on the mass pan and the gimbal system above the mass pan is shown. Bottom left: LEGO T-block serving as the central pivot with balls and V-blocks for kinematic realignment. An identical set exists on the opposing face of the balance.}
\label{fig:mosaic}
\end{figure}

The magnet system we chose to generate this radial magnetic field consists of a pair of neodymium (N48) ring magnets, one pair per coil. For simplicity, we recommend keeping the system an open-field design, i.e., ``yokeless,'' meaning no additional ferromagnetic material to guide the magnetic flux direction. The dimensions of the ring magnets were chosen such that they could fit inside the PVC pipe coil former with approximately 0.5\,cm clearance all around. A brass threaded rod secured to a non-magnetic base plate (wood, aluminum, etc.)  provides the vertical guide for each magnet system (see Fig.~\ref{fig:mosaic}). The magnets are oriented on the brass rod such that they repel each other, and two aluminum nuts on either side of the magnets constrain the repulsive force, also setting their separation distance. This design allows us to adjust the distance between the magnets and the geometrical center of the magnet assembly. 
 
Each coil former was made from a standard 1-inch PVC water pipe with end caps glued to it. Any nonmagnetic, rigid, cylindrical body will suffice in serving as the coil former. The coil was manually wound onto the PVC pipe using a very low-speed lathe spindle and each layer of wire was potted with spray glue. We chose to use AWG-36 wire with about 3000 windings. In our system, a current of 2.7\;mA generated about 0.1\;N of force. The total resistance of the wire was 450\;$\Omega$.

The coils can be constructed without a lathe by either hand-winding or by using a battery-powered drill. Using a lathe to turn down the PVC pipe is an optional step, which we chose to use because it allowed a deeper groove for more windings. Increasing the number of windings on the coil increases the vertical electromagnetic force generated, hence increasing the $BL$ factor.

A small hole was drilled into each end cap top where a LEGO cross axle was attached vertically, allowing each to hang rigidly beneath their corresponding mass pan (see Fig.~\ref{fig:mosaic}). The mass pan was suspended from three rigid rods linking to a LEGO universal joint (part no. 61903). This dual-gimbal system hangs from a set of two freely pivoting axles parallel to the central pivot (part no.\ 4208204) connecting to the balance arm. The central pivot (T-brick part no.\ 4211713) has a ``knife edge'' radius of approximately 3.1\;mm and rests on a smooth surface.

The whole balance measures approximately 43~cm $\times$ 36~cm $\times$ 10~cm and has a mass of 4~kg, including the wooden base board. 

\section{Electronics and Data Acquisition}

We employ two USB devices, a U6 from Labjack and a 1002\_0 from Phidget, to connect the LEGO watt balance to a laptop computer. The U6 is used to measure the position of the balance beam, the induced voltage, and the current in each coil. We connect a sixth input to a LEGO handheld controller (potentiometer) that allows students to manually tare the balance, providing an interactive element at science fairs and demonstrations. The 1002\_0 is a four-channel analog output that is capable of producing a voltage between $-10$\;V and $+10$\;V. Each channel can source up to 20\;mA. One channel is used for each coil. One channel is connected to a double-throw, double-pole relay. This relay allows the analog output to disconnect from either coil. One coil serves as the sine-driven actuator while the induced voltage can be measured in the other. The relay toggles between the two coils, allowing the operator to select which one is the driver. The last output channel is used to remove the bias voltage in the photodiode, as explained below in Sec.~V. This allows the use of a smaller gain setting on the analog-to-digital converter that reads the photodiode.

\begin{figure}[h!]
\includegraphics[width=8.5cm]{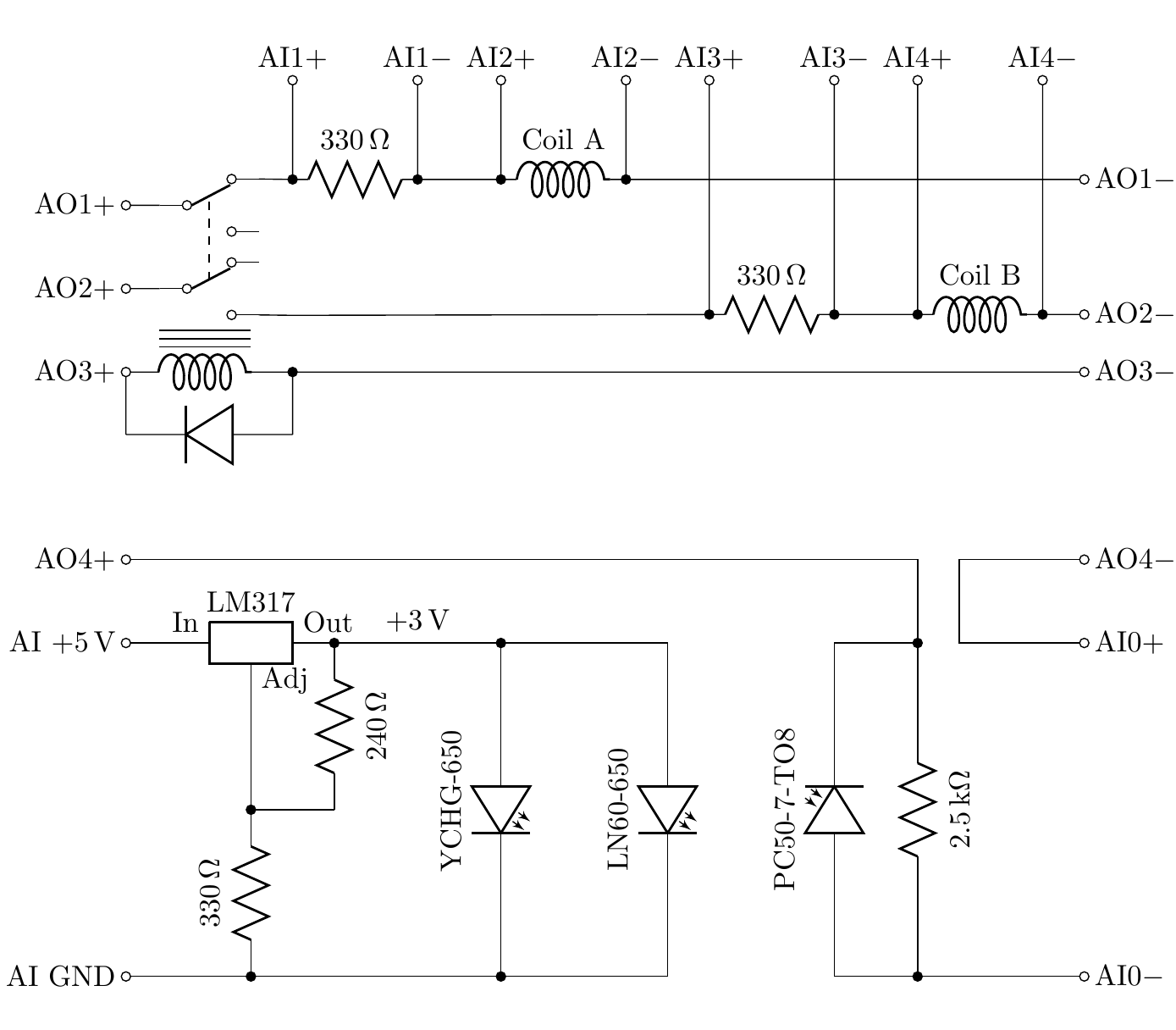}
\caption{Circuit diagram for the LEGO watt balance. The top diagram connects one of the two coils to the analog output via a double pole relay. The bottom diagram shows the power supply for the two laser pointers and the photodiode.}
\label{fig:cdiagram}
\end{figure}

We designed the circuit to keep the part count low (seven resistors, one relay, and one voltage regulator), to allow for easy construction. Figure~\ref{fig:cdiagram} shows the circuit diagram. The top circuit is used to measure the induced voltage and current in each coil. 

The circuit on the bottom left provides the 3\;V for the two laser diodes (see Sec.~V for functions of the optical system). The circuit on the right in the diagram reads the position in the following way: The photo current produced in the photodiode is proportional to the balance position. The photocurrent flows through R5, the 2.5\;k$\Omega$ resistor. The voltage drop across R5 is added to the analog output voltage produced by AO4 and the sum is measured. By setting AO4 negative, 0\;V can be obtained when the balance is at the nominal weighing position.

\begin{figure}[h!]
\includegraphics[width=8.5cm]{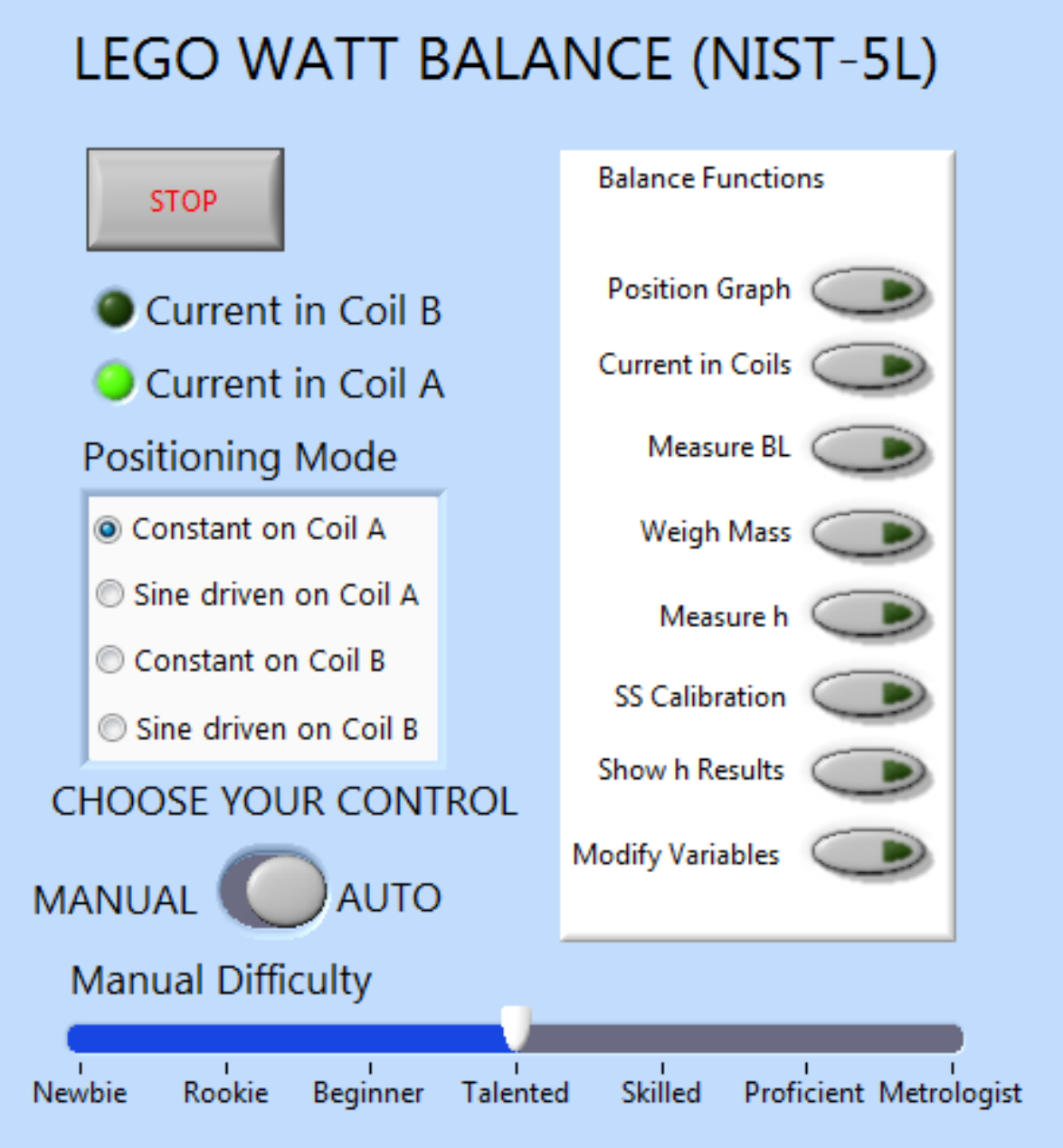}
\caption{The front panel of the LEGO watt balance control system. It allows the operator to calibrate the system, weigh small masses, and measure Planck's constant. Weighing mode (or force mode) can be done either automatically or manually.}
\label{fig:labview}
\end{figure}

A custom executable program has been designed to control the LEGO watt balance. If interested in obtaining the free executable and CAD file, please visit the American Journal of Physics Electronics Archive found in Ref.~\onlinecite{software}. Figure~\ref{fig:labview} shows a screenshot of the main interface.

\section{Measurement}

Like a professional watt balance, the LEGO watt balance must undergo a series of alignments and calibrations prior to starting the experiment, detailed in the following four-step procedure. It is important to calibrate and sense the balance's angular  position. Again, there are many ways to achieve this, but here the angular position of the balance was monitored using a shadow sensor. The system consists of a laser pointer and a photodiode near the lower edge of one arm of the balance. When the balance moves, it gradually obstructs the optical path of the laser, thereby changing the intensity of light hitting the photo detector. A second laser pointer mounted on top of the balance serves as an optical lever for calibrating the shadow sensor, as we will describe shortly. 

Once these prerequisites have been achieved, a complete determination of a mass or the value of the Planck constant can begin using a common A-B-A measurement technique. This repetition method is used to cancel the time-dependent drift associated with measurements. For instance, one can interleave velocity mode, then force mode, then velocity mode again. Ideally, these measurements are done such that the instrument undergoes as little change as possible, or by performing the measurements in quick succession, neither moving nor tinkering with the balance in between measurements. Once the system is properly aligned and calibrated, a full determination of $h$ through measuring $\BLV$  in velocity mode and $\BLF$ in force mode is possible. For reference, our experienced operators could perform the following alignment, calibration, and measurement procedure in about 10 minutes.

\subsection{Calibrating the Shadow Sensor}

If a shadow sensor and optical lever are indeed chosen for position sensing, a four-step process is advised to prepare the balance for calibration.

\begin{enumerate}
\item Place the LEGO watt balance on a flat, level surface located a few meters from a wall or vertical structure. 

\item Shine the laser pointer mounted on top of the balance at a wall a few meters away as in Fig.~\ref{fig:mosaic}. Ideally, a ruler or grid paper is taped to the wall where the laser spot is located. Measure the distance $d$ from the pivot point of the balance to the wall. 

\item Align the balance beam to its support tower. Ensure that the balance is not rotated around the $y$ and $z$ axes (the coordinate system is shown in Fig.~\ref{fig:app}). Looking from the top, the clearances between the beam and the support tower have to be evenly spaced on each side. Our version of the balance has several auxiliary parts, i.e., balls that engage in V grooves and a swivel bracket that aid in the balance alignment. However, it is also possible to perform alignment without these parts. Also, it is good practice to check if the balance is fairly leveled when absent of masses.

\item Concentrically align each magnet system to its corresponding coil. Each magnet system is mounted on X-Y adjustable plates that may slide around until concentricity is reached. Each plate can be clamped down afterwards. It is important to ensure that the coils are not touching the magnets. 
\end{enumerate}

After these four alignment steps, the instrument is ready for calibration. The balance is servoed to a few different angles, which causes the shadow sensor to detect differing light intensities and convert them into voltages $V_i$. For each voltage, the position $x_i$ of the light spot on the ruler is measured. In addition to these points, we note the position $x_0$ of the light when the balance is horizontal. The balance angle is then determined as $\theta_i = (x_i-x_0)/d$ and the coil height is calculated by multiplying the balance angle by the effective radius, or $z_i=r_\mathrm{eff} \theta_i$. The effective radius is found by measuring the distance from the knife edge to the mass pan universal joint. For the balance described here, $r_\mathrm{eff} = 175$\;mm. 

The optical sensing method described above was contrived to drive the measurement uncertainty down to reach our 1\% goal. If this goal is not required, easier methods can be used, i.e., directly measuring the coil height for differing servoed positions.

\begin{figure}[ht!]
\includegraphics[width=8.5cm]{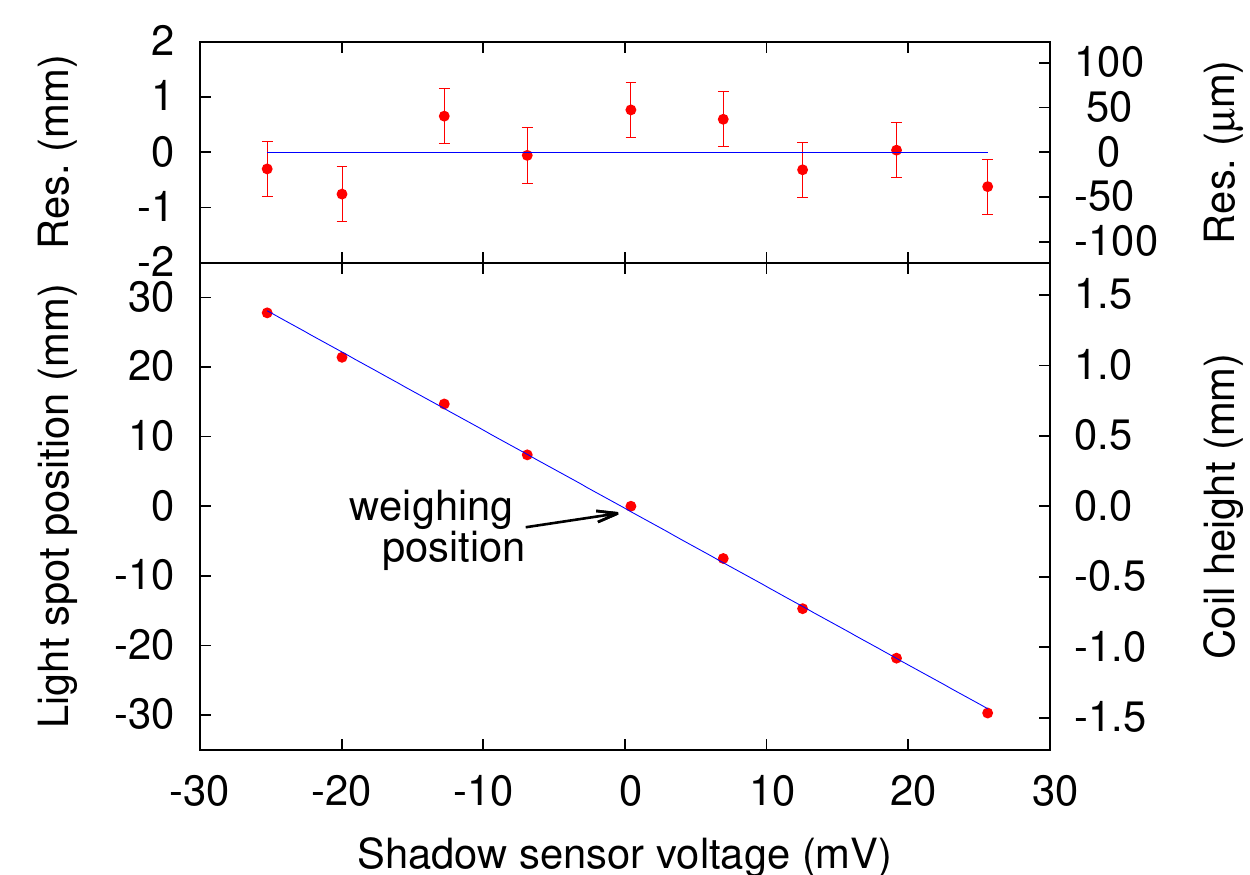}
\caption{Calibration of the shadow sensor. The balance is servoed to 9 different shadow sensor voltages. For each voltage, the position of the light spot on the wall, in this case 3489\,mm away, is measured. The relationship between the position of the light spot on the wall and the shadow sensor voltage is almost linear. The solid line indicates the best fit to the data. The upper graph shows the residuals between the fit line and the measured data points. We attributed an uncertainty of 0.5\, mm (represented by the error bars) to the position measurement of the light spot. Judging from the residuals, that seems reasonable.}
\label{fig:cal_data}
\end{figure}

Within a reasonable range, the voltage produced by the shadow sensor is a linear function of the coil height. Hence the coil height can be obtained as  $z(V) = b (V-V_o)$.  A best-fit line to the data $(z_i,V_i)$ yields $b$ and $V_o$. Figure~\ref{fig:cal_data} shows an example of such a calibration. 

\subsection{Velocity Mode Measurement}

As stated before, velocity mode measurement ($\BLV$) is the key for characterizing the electromagnetic properties of the balance, and is the first measurement step toward obtaining an $h$ or mass value. Our chosen method was to use the information from our calibrated optical displacement sensor and simply take its time derivative to calculate velocity.

If one wants to perform a watt balance experiment using Coil A, then Coil B will be used to drive the balance in a sinusoidal motion; see Fig.~\ref{fig:app}. Again, there are many ways to actuate velocity mode. We chose a symmetric design such that either arm could be the driver, but other ideas such as a LEGO miniature piston engine have also been tried.~\cite{Quinn13} Because we arbitrarily chose Coil A as the measurement coil and Coil B as the driver, we will continue this nomenclature for consistency and clarity. Using the language of control theory, Coil B was the input driven with an open-loop sinusoidal voltage, and the output balance position was detected by the shadow sensor.

A sinusoidal driving signal resulting in a 1\;mm coil displacement and a period of 1.5\;s  seemed to be a good starting point for our balance. We sampled the Labjack analog input device at a rate of $\Delta=1$\;ms and obtained values for the induced voltage on Coil A, $V(i\Delta)$, and the shadow sensor voltage $V_{SS}(i\Delta)$, where $i$ is the sample number. The coil position $z(i\Delta)$ was then extracted from the shadow sensor voltage. The sampled data were filtered and the coil velocity was obtained by taking the numerical derivative:
\begin{equation}
v(i\Delta) = \frac{z\big( (i+1)\Delta\big)-z\big( (i-1)\Delta\big)}{2\Delta}.
\end{equation}

For the pairs of voltages and velocities, a best-fit line was calculated whose slope was $\BLV$. For simplicity, we assumed that $\BLV$ did not vary significantly along the coil's trajectory. Since the coil moved only 2\;mm, this seems like a reasonable assumption.

\begin{figure}[ht!]
\includegraphics[width=8.5cm]{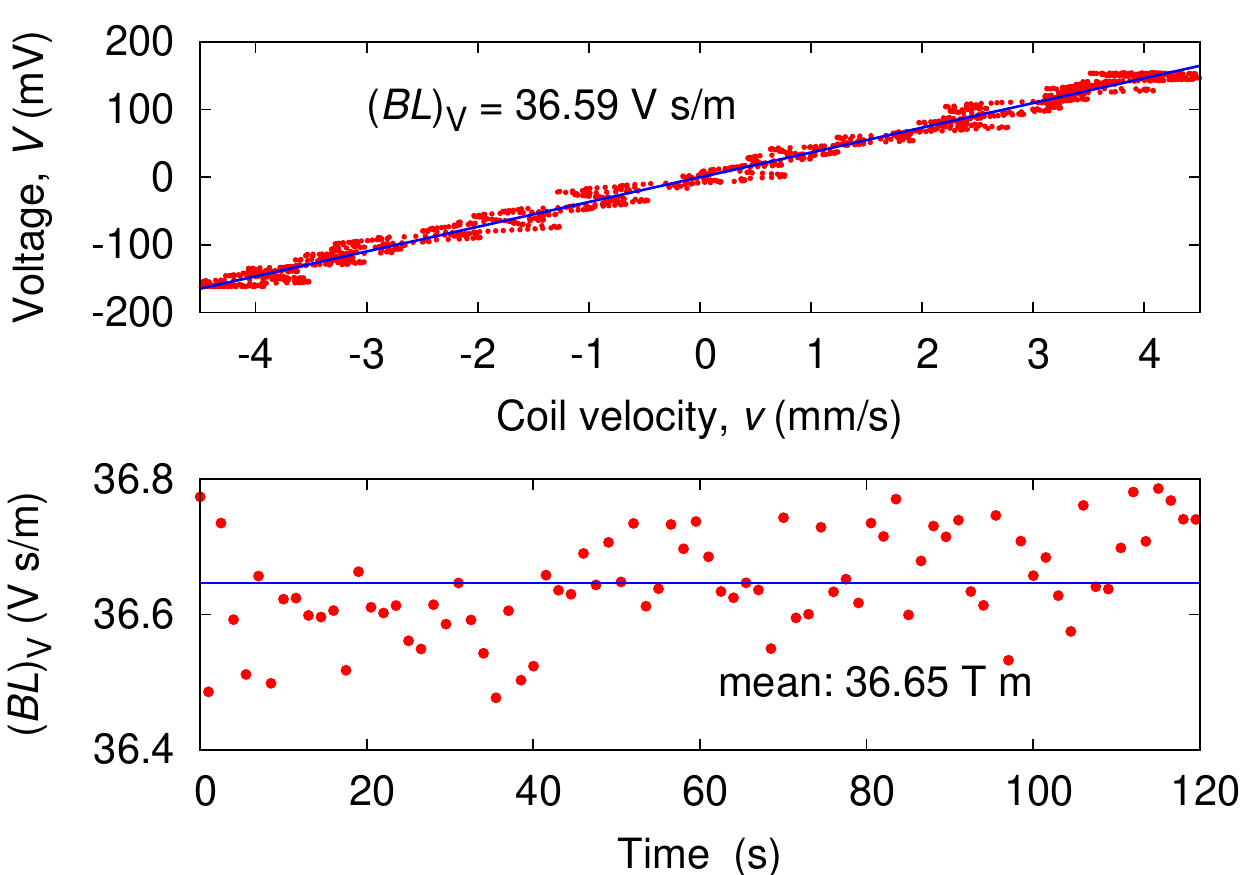}
\caption{The top graph shows the measured voltages and velocities of the coil for one period, i.e., 1.5\,s. The slope of the solid line which is the best linear fit to the data gives the measured flux density, $BL$. The bottom graph shows the result of 80 such determinations. The relative standard deviation of the data is 0.2\,\%. For a possible explanation of the small drift see the main text.}
\label{fig:velomode}
\end{figure}

The top graph in Fig.~\ref{fig:velomode} shows the measured values of the induced voltage versus the calculated coil velocity. The data shown were taken during  one period of a sinusoidal motion of the coil. The slope\cite{Note3} is 36.59\;V\,s/m. The bottom graph shows 80 of these measurements for a total of 120\;s. A value of the flux integral is determined every motion period. From this data we obtained a mean value
\begin{equation}
\BLV = 36.65\;\textrm{V\,s/m}.
\end{equation}

 The relative standard deviation of the data was 0.21\,\%. A small relative drift of $3\times 10^{-5}$\;s$^{-1}$ was apparent in the data. This drift can be explained by small temperature changes of the magnet. The remanent field of NdFeB magnet changes relatively by about $-1\times 10^{-3}\;$K$^{-1}$. Hence, a temperature change of -0.01\;K/s would explain the observed drift in the $\BLV$. Here, we ignore the observed drift of the $\BLV$ and assign the mean value.

\subsection{Force Mode Measurement}
Coil A is used in the force mode to apply an electromagnetic force to one arm of the balance. The force is easily created by running a current through the coil, but keep in mind that the magnitude must be controlled somehow to hold the balance in its null position after masses are added or removed from the mass pans. The most direct way to control the current is to simply watch the balance and manually adjust the magnitude of the current until balance is restored. This option is available using the LEGO potentiometer. Simply connect the potentiometer to the coil in series with a battery to form a closed circuit. The projected laser spot on the wall, used to calibrate the shadow sensor, can be used as a target for restoring the balance by manually adjusting the potentiometer.

For users more familiar with control theory and application, the manual feedback can be automated to achieve more consistent results. For instance, the output position of the balance can be detected by the shadow sensor and employed as the control variable for an analog or digital controller. The measured position is then continuously compared by the controller to a desired position, or setpoint (typically a null position), and the error between the two used to continuously update the current input to Coil~A. In our system, a digital feedback control software tool operates on the data acquisition and control laptop. The controller generates currents that are proportional to the measured error and the integral and derivative of this value with respect to time. Such a scheme is referred to as PID control, where the acronym stands for proportional, integral, and derivative control.

\begin{figure}[ht!]
\includegraphics[width=8.5cm]{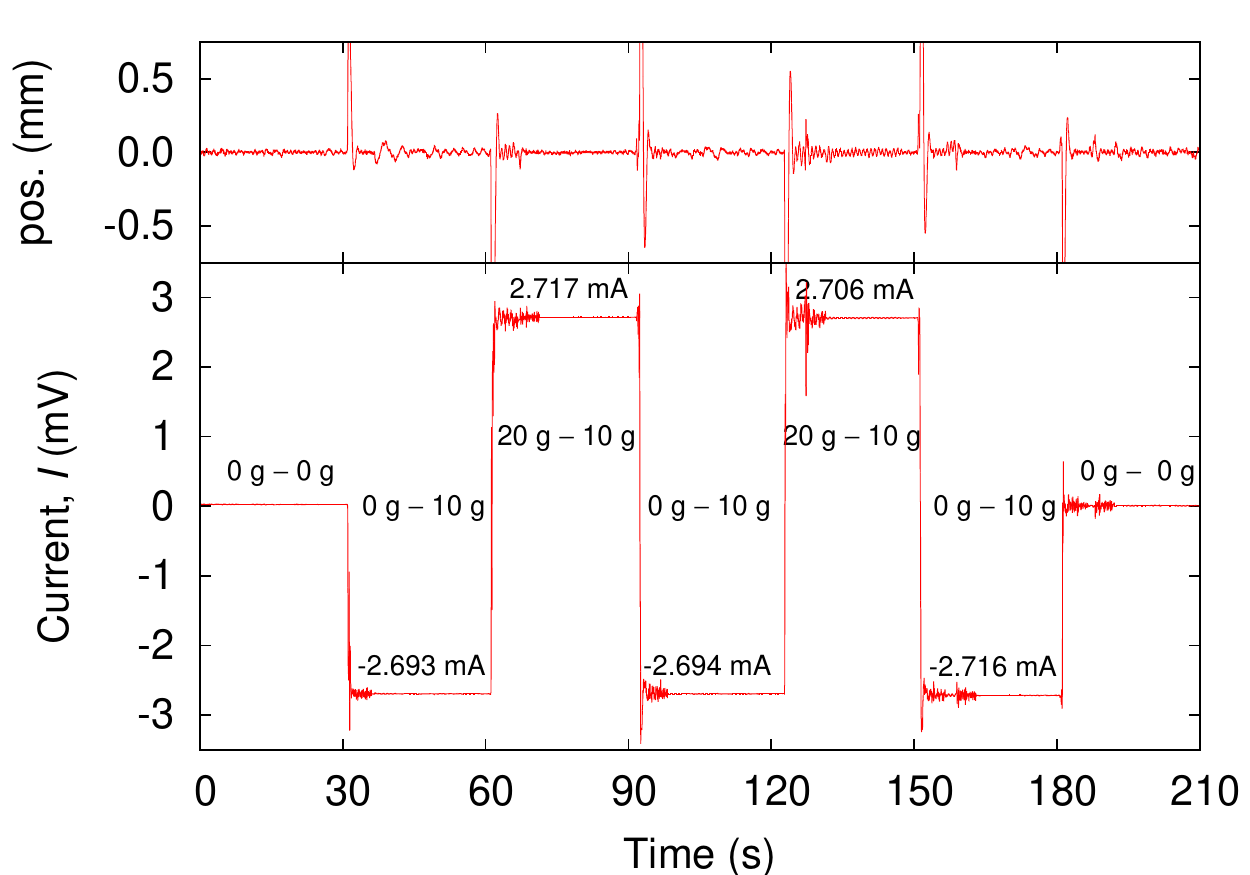}
\caption{Force mode in the time domain. The lower graph shows the current required to maintain the balance at a nominal position for seven different load states. The load states are abbreviated by differences.  The minuend denotes the mass on the mass pan above Coil A and the subtrahend the mass on the mass pan above Coil B. The mass difference multiplied by the local gravitational acceleration is equal to the force produced by the coil. The software PID controller that is used to servo the balance employs two different gain settings. The change in noise in the measured current occurs when the gain is switched. The top graph shows the position of the coil as a proxy of the balance angle. Adding and removing a mass leads to a spike in position up to 2\,mm. The servo quickly reestablishes the nominal weighing position.}
\label{fig:forcemode}
\end{figure}

Figure~\ref{fig:forcemode} shows the measurement sequence in the force mode. In this example, the measurement was performed in seven steps, each lasting 30\;s. The steps were:

\begin{enumerate}
\item Both balance pans are empty and the current required to hold the balance at its weighing position is small. 
\item A tare mass $m_T=10$\;g  is added to the pan above Coil B. The exact mass is irrelevant as it will drop out in the final equation. The current $I_1=-2.693$\;mA is necessary to maintain the balance position. The current is given by
\begin{equation}
I_1 \BLF = -m_T g.
\label{eq:I1}
\end{equation}
\item The calibrated mass, here $m=20.2$\;g, is added to the pan above Coil~A. This time a positive current, $I_2=2.717$\;mA, is required to servo the balance. The equation 
\begin{equation}
I_2 \BLF -m g = -m_T g
\label{eq:I2}
\end{equation}
describes this weighing. Subtracting  Eq.~\ref{eq:I2} from Eq.~\ref{eq:I1} is sufficient to get an estimate of $\BLF$,
\begin{equation}
m g =(I_2-I_1) \BLF \Longrightarrow \BLF=\frac{mg}{I_2-I_1} .
\label{eq:I2m2}
\end{equation}
However, to cancel out drift and to get an idea how big the drift is, it is always a good idea to perform a couple more weighings.
\item Another weighing with the Coil A calibrated mass removed determines
\begin{equation}
I_3 \BLF = -m_T g.
\end{equation}
\item A second weighing with the Coil A calibrated mass added to the pan yields
\begin{equation}
I_4 \BLF -m g = -m_T g.
\end{equation}
\item A third weighing with the Coil A calibrated mass removed yields
\begin{equation}
I_5 \BLF = -m_T g.
\end{equation}
\item We finally remove both masses and check if the balance is back at the nominal position and if the current to servo the balance with no weights on either pan has remained stable.
\end{enumerate}

Using the above observations, the following number can be calculated:\cite{Swanson10}
\begin{equation}
I=-\frac{1}{3} \left( I_1 + I_3 + I_5 \right) +\frac{1}{2} \left( I_2 + I_4 \right) = \frac{mg}{\BLF}.
\end{equation}

In order to obtain the flux integral from the force mode, one needs the local gravitational acceleration~$g$. Your local gravitational acceleration can be obtained from a website provided by the National Oceanic and Atmospheric Administration.\cite{noaa} For our geographical coordinates at NIST Gaithersburg
(Latt: $39.1261^{\circ}$N, Long: $77.2211^{\circ}$W, Elevation: 124.304\,m),
the website yielded $g=9.80103\;$m/s$^2$ with a relative uncertainty of $2\times 10^{-6}$. This uncertainty was well below what we needed for a 1\%-level measurement. 

With the above numbers and $m=20.2\;$g, we obtain
\begin{equation}
\BLF = \frac{0.0202\;\mbox{kg}  \cdot 9.80103\;\mbox{m}/\mbox{s}^2}{0.0054125\;\mbox{A}} = 36.58\,\mbox{N}/\mbox{A}
\end{equation}

\subsection{A value for $h$}

To obtain a value for $h$, the ratio of the flux integrals obtained in force mode and velocity mode was multiplied with $h_{90}$ as described in Eq.~\ref{eq:bl_comb}. Here, we obtained
\begin{equation}
h = h_{90} \frac{36.58\;\mbox{N}/\mbox{A}}{36.65\;\mbox{Vs}/\mbox{m} } = 0.998\; h_{90}= 6.61\times 10^{-34}\;\mbox{J\,s}.
\end{equation}

Determining a value for the Planck constant is half the work; the other half is to determine the measurement's uncertainty. We believe that a measurement of the Planck constant with a relative uncertainty of 1\% or less is possible with this LEGO watt balance. 

For example, a source of uncertainty that is easy to understand comes from the distance measurement in the shadow sensor calibration. If the distance from the laser diode to the wall is measured with a measuring tape to be 3000\;mm with an uncertainty of 3\;mm, the relative uncertainty associated with this would be 0.1\%. Also, if the laser spot oscillates by $\pm30$\;mm and the vertical wall ruler can be read with an uncertainty  of 0.5\;mm, then the relative uncertainty of this measurement is 0.83\%, and this is clearly the dominant source of uncertainty in this measurement.  A good metrologist will identify the largest sources of uncertainty and will try to reduce these.

Large contributors to bias and measurement uncertainty are offset forces produced in the large-radius knife edge, parasitic motions of the coil during velocity mode, and horizontal forces in force mode which arise from misalignments. An uncertainty analysis is beyond the scope of this article and we leave this exercise to the interested reader. Several inspiring articles can be found in the literature that provide details on how to assess these uncertainties, e.g., Refs.\ \onlinecite{Steiner06, Robinson12, Kibble14}.

\section{Summary}

\begin{figure}[h!]
\includegraphics[width=8.5cm]{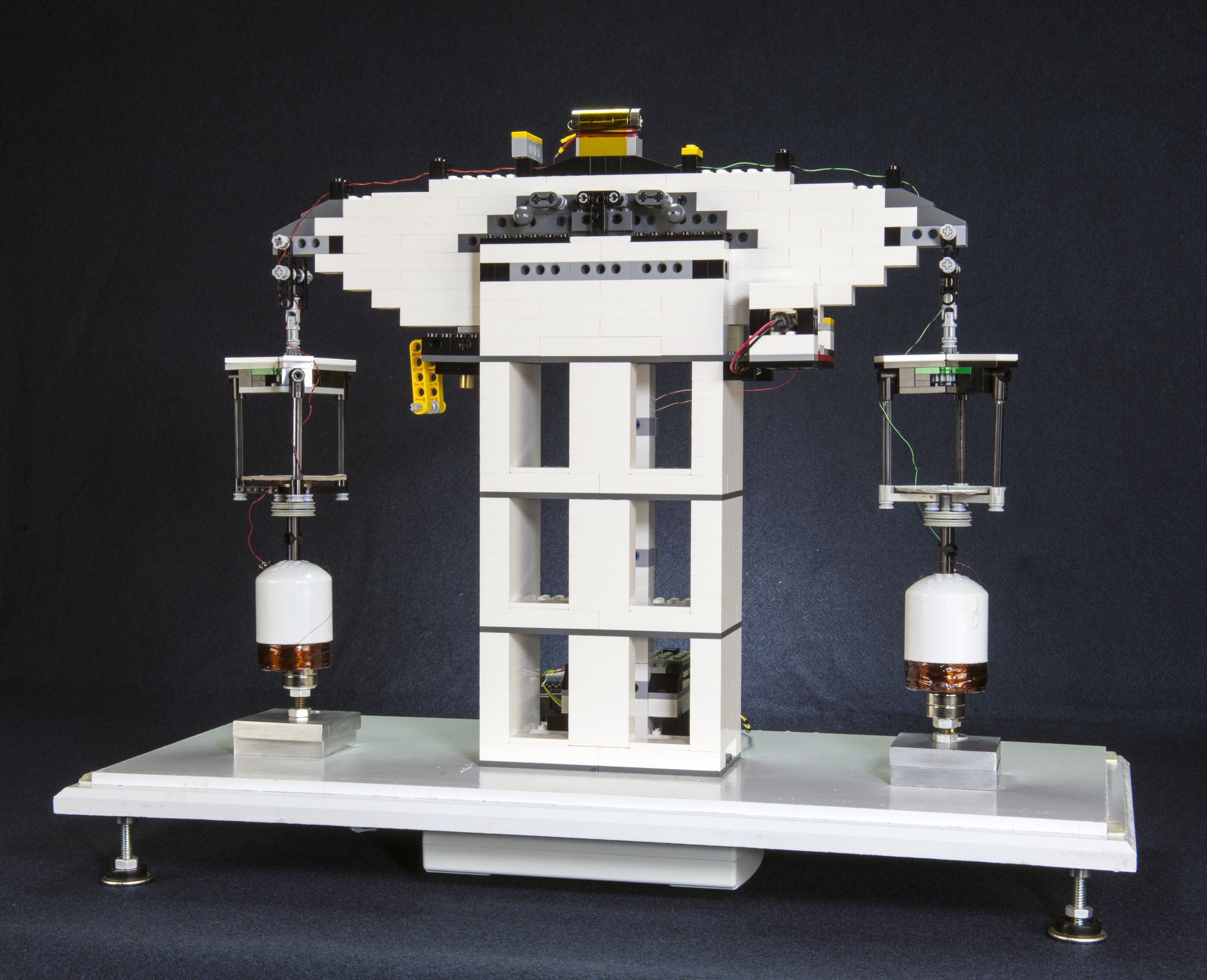}
\caption{A photograph of our first prototype LEGO watt balance. (The parts list in Appendix~A reflects later models.)}
\label{fig:photo}
\end{figure}

In 2013 and 2014 we built five LEGO watt balances with our original prototype shown in Fig.~\ref{fig:photo}. These balances were demonstrated and received with enthusiastic responses in science fairs, classrooms, and with visitors coming to NIST. This success prompted us to write this article to promote building these devices for STEM education. What will building and operating such a device accomplish? 

In the new SI, the kilogram will be defined via fixing the Planck constant,~\cite{Knotts14} through a procedure that is not apparent to most people.  One can show how it is possible to generate a mechanical force, whose value is precisely given by electrical measurements. Unfortunately, it still requires some abstraction to explain how electrical power is related to the Planck constant via the Josephson effect and the quantum Hall effect. But once that bridge is crossed, the relationship between mass and $h$ can be made clear. From there, it is fascinating to ponder the implications of the redefinition of the kilogram: the Planck constant, a natural constant found in quantum mechanics, can be directly used to determine mass on a macroscopic scale---or at any scale, for that matter.  With this LEGO balance, one can determine the absolute mass of a small object without needing any comparison or traceability to reference mass standards. It is also interesting to consider how an apparatus assembled from plastic bricks can measure $h$ in a classroom or living room setting with an uncertainty of only 1\%.

Besides these top-level concepts, we found that the LEGO watt balance provides ample teachable moments. For example, our balance controls included a ``manual feedback'' option where an operator could rotate a potentiometer to try and null the balance. Most people who find it hard to control the balance are amazed to see how effortlessly and precisely a PID controller achieves the task. This provides a nice segue into control theory.

Closer to home, questions at the heart of metrology arise while constructing such an experiment:  How does one measure something?  Are all measurements comparisons?  What are accuracy and precision? What is the error in this measurement?  Answering these questions provides an opportunity to teach the audience about the importance of measurements for society. Getting the audience interested in metrology is the intent of the LEGO watt balance. This goal has been achieved every time we demonstrated our glorified toy.

As an additional outreach tool we have created a Facebook page\cite{facebook} showcasing diagrams and photographs of the LEGO watt balance. An instructional video is under development and will be posted there when it is ready. The goal of the page is to cultivate a community for enthusiasts to share their building experiences and exchange insights. It is our dream to see others construct their own instruments, improve on our design, and even surpass 1\% relative uncertainty. Don't forget to ``like'' our page!

In conclusion, the LEGO watt balance combines three important ingredients:  science, technology, and fun.

\appendix
\section{List of parts}

The table below shows the majority of the components we used to build our LEGO watt balance. Each LEGO engineer is encouraged to explore other building components to create a more optimized and personalized instrument. The prices are accurate as of 2014 and do not include shipping and handling. Although we spent over \$600 on this project, it can be built for significantly less. For example, we also chose to employ the National Instruments USB-6001 Data Acquisition (DAQ) device (\$189), which replaces both the Labjack DAQ and the Phidget Analog 4 Output, reducing the total cost by \$200. We have verified its functionality and the corresponding circuit diagram and software are included in the electronic supplement.\cite{software} 

In addition to the parts listed below, a wooden base, wires to connect the electrical circuits, and a spool of wire to wind the coil are required and can be purchased from a variety of vendors.

\newpage
\begin{longtable*}{llrr}
\centering
\small
	&		&		&		\\
\textbf{Part Name}	&	\textbf{Part No.}	&	\textbf{Quantity}	&	\textbf{Total Price (\$)}	\\
\endhead
	&		&		&		\\

Custom LEGO Watt Balance Software &  & 1 & Free\\

Brick 2x4 	&	300101	&	75	&	22.50	\\
Brick 2x8 	&	6033776	&	75	&	37.50	\\
Brick 1x2 with cross hole 	&	4233487	&	6	&	2.10	\\
T-Beam 3x3 w/hole 	&	4552347	&	2	&	0.60	\\
Technic Brick 1x2	&	370026	&	18	&	2.70	\\
Technic Brick 1x4 	&	4211441	&	66	&	16.50	\\
Technic Brick 1x5 Thin	&	32017	&	4	&	0.80	\\
Plate 8x8  	&	4210802	&	9	&	9.90	\\
Plate 1x2  	&	4211398	&	14	&	1.40	\\
Plate 1x4  	&	4211445	&	3	&	0.45	\\
Plate 2x3  	&	4211396	&	6	&	1.20	\\
Cross Axle 2M w/ Groove	&	4109810	&	8	&	0.80	\\
Cross Axle 3M   & 4211815 & 2 & 0.20  \\
Cross Axle 5M  	&	4211639	&	6	&	1.20	\\
Cross Axle 8M 	&	370726	&	8	&	1.60	\\
Bush for Cross Axle 	&	4211622	&	14	&	2.10	\\
1/2 Bush for Cross Axle 	&	4211573	&	32	&	3.20	\\
Double Bush 3M	&	4560175	&	4	&	0.80	\\
Roof Tile 2x2/45 deg Inv.	&	366026	&	2	&	0.40	\\
Roof Tile 2x3/25 deg 	&	4211106	&	6	&	1.20	\\
Roof Tile 2X3/25 deg Inv.	&	374726	&	4	&	0.80	\\
Connector Peg W. Friction 3M	&	4514553	&	8	&	2.00	\\
Connector Peg/Cross Axle	&	4666579	&	6	&	0.60	\\
Catch w. Cross Hole 	&	4107081	&	8	&	1.60	\\
Flat Tile 2x4 	&	4560178	&	2	&	0.60	\\
Hinge 1x2 Lower Part 	&	383101	&	6	&	1.50	\\
Hinge 1x2 Upper Part 	&	6011456	&	6	&	1.50	\\
Double Conical Wheel Z12 1M	&	4177431	&	4	&	1.20	\\
Angle Element, 180 Degrees [2]	&	4107783	&	2	&	0.40	\\
Technic Beam 1 x 4 x 0.5 with Boss  	&	2825 / 32006	&	6	&	0.30	\\
Technic Beam 2 Beam w/ Angled Ball Joint 	&	50923 / 59141	&	2	&	0.13	\\
Wedge Belt Wheel 	&	2786 / 4185	&	4	&	1.00	\\
Gear with 8 Teeth (Narrow) 	&	3647	&	2	&	0.20	\\
Universal Joint	&	61903	&	2	&	0.94	\\

Multifunction DAQ with USB - 16 Bit	&	U6	&	1	&	299.00	\\

PhidgetAnalog 4 Output	&	1002\_0	&	1	&	90.00	\\

Focus Line Red Laser Module \textless 1mW	&	YCHG-650	&	1	&	15.00	\\
Line Laser Module (650nm)  \textless 1mW	&	LN60-650	&	1	&	15.00	\\

Photodiode 7.98mm Dia Area	&	718-PC50-7-TO8	&	1	&	61.63	\\
Low signal Relay	&	769-TXS2-4.5V	&	1	&	4.58	\\
Resistors 240 Ohms &	291-240-RC	&	1	&	0.10	\\
Resistors 330 Ohms &	291-330-RC	&	4	&	0.40	\\
Resistors 1500 Ohms &	291-1.5k-RC	&	1	&	0.10	\\
Linear Voltage Regulator &	511-LM317T	&	1	&	0.72	\\

N48 grade - 3/4(OD) x 1/4(ID) x 1/2 in. ring magnet	&	NR011-0	&	4	&	15.96	\\

Brass Threaded Rod - 1/4"-20 Thread, 1' length	&	98812A039	&	1	&	2.65	\\
White PVC Pipe Fitting	&	4880K53	&	2	&	1.00	\\
White PVC Unthreaded Pipe	&	48925K93	&	1	&	5.27	\\

	&		&		&		\\
\multicolumn{3}{l}{Total} & 632.47	\\

\end{longtable*}

\section{An alternative circuit  for the NIST LEGO Watt Balance}
We built our first LEGO Watt Balance in the summer of 2012. Since then we have built four more devices, iteratively streamlining various components. In the end, we found a cheaper and easier way to build the instrument by using two simplifications:

1) National Instruments\footnote{Certain commercial equipment, instruments, or materials are identified in this paper in order to specify the experimental procedure adequately. Such identification is not intended to imply recommendation or endorsement by the National Institute of Standards and Technology, nor is it intended to imply that the materials or equipment identified are necessarily the best available for the purpose.} released a new USB data acquisition device, the USB-6001. This device has bipolar analog output, which is necessary for the LEGO Watt Balance. This single device replaces both the Labjack and Phidget described in the text. 

2) We also discovered the laser modules can be powered with 5\,V. Hence, there is no need to use a voltage regulator.

\begin{figure*}[t!]
\centering
\includegraphics[width=6in]{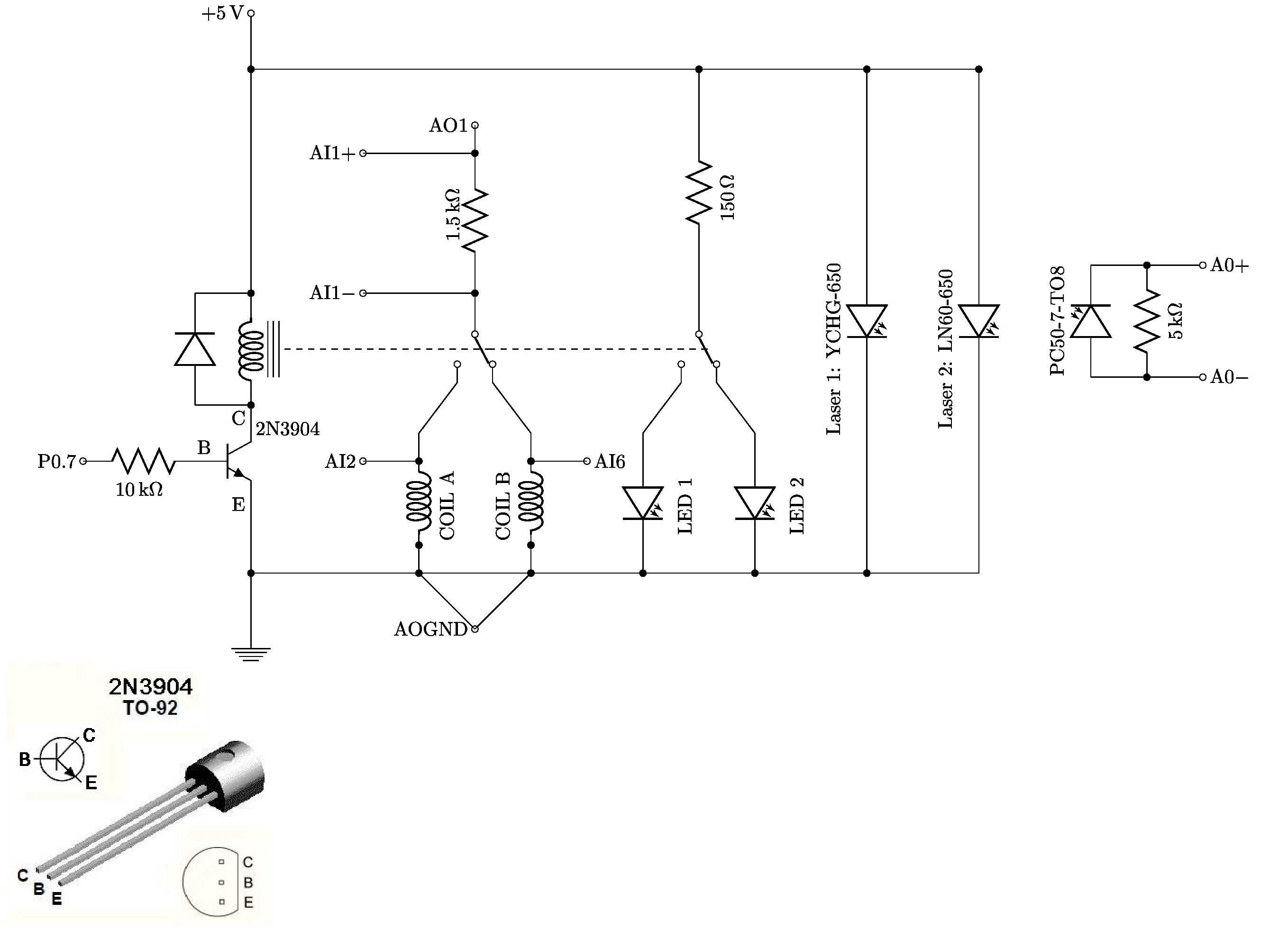}
\caption{The simplified version of the watt balance circuit.}
\label{fig:simplecircuit}
\end{figure*}

The new circuit diagram is shown in Fig.~\ref{fig:simplecircuit}. We will provide software for both versions of the electronics.


\begin{thebibliography}{10}

\bibitem{realization}
Joint Committee for Guides in Metrology, Working Group 2, \textit{International vocabulary of metrology---Basic and general concepts and associated terms (VIM)}, 3rd edition (JCGM, 2008), p.~46.

\bibitem{Seifert14}
F.~Seifert, A.~Panna, S.~Li, B.~Han, L.~Chao, A.~Cao, D.~Haddad, H.~Choi,
  L.~Haley, and S.~Schlamminger, ``{Construction, Measurement, Shimming, and
  Performance of the NIST-4 Magnet System},'' IEEE
  Trans. Instrum. Meas. \textbf{63}, 3027--3038, (2014). 
  
\bibitem{Stock13}
M.~Stock, ``{Watt balance experiments for the determination of the Planck constant and the redefinition of the kilogram},'' {Metrologia} \textbf{50}, R1--R16 (2013).

\bibitem{Kibble75}
B.~Kibble, ``{A Measurement of the Gyromagnetic Ratio of the Proton by the
  Strong Field Method},'' in \textit{Atomic Masses and Fundamental Constants}, vol. 5,
  edited by J. H. Sanders and A. H. Wapstra (Plenum, New York, 1976), pp.\ 545--551.

\bibitem{Quinn13}
T.~Quinn, L.~Quinn, and R.~Davis, ``{A simple watt balance for the absolute
  determination of mass},'' {Physics Education} \textbf{48}, 601--606
 (2013).


\bibitem{Note1}
Certain commercial equipment, instruments, and materials are identified in this
  paper in order to specify the experimental procedure adequately. Such
  identification is not intended to imply recommendation or endorsement by the
  National Institute of Standards and Technology, nor is it intended to imply
  that the materials or equipment identified are necessarily the best available
  for the purpose.

\bibitem{Clarke70}
J.~Clarke, ``{The Josephson Effect and e/h},'' {Am. J. Phys.} \textbf{38}, 1071--1095 (1970).

\bibitem{Tang12}
Y.~Tang, V.~N. Ojha, S.~Schlamminger, A.~R\"{u}fenacht, C.~Burroughs,
  P.~Dresselhaus, and S.~Benz, ``{A 10V programmable Josephson voltage standard
  and its applications for voltage metrology},'' {Metrologia} \textbf{49}, 635--643 (2012).

\bibitem{Eisenstein92}
J.~Eisenstein, ``{The quantum Hall effect},'' {Am. J. Phys.} \textbf{61}, 179--183 (1993).

\bibitem{Rand01}
R.~E. Elmquist, M.~E. Cage, Y.-h. Tang, A.-M. Jeffery, J.~R.~J. Kinard, R.~F.
  Dziuba, N.~M. Dziuba, and E.~R. Williams, ``{The Ampere and Electrical
  Standards},'' {J. Res. Natl. Inst. Stand. Technol.} \textbf{106},
  65--103 (2001).

\bibitem{Quinn89}
T.~Quinn, ``{News from the BIPM},'' {Metrologia} \textbf{26}, 
  69--74 (1989).

\bibitem{Taylor89}
B.~Taylor and T.~Witt, ``{New international electrical reference standards
  based on the Josephson and quantum Hall effects},'' {Metrologia}
  \textbf{26}, 47--62 (1989).

\bibitem{noaa} National Oceanic and Atmospheric Administration,
``{Surface Gravity Prediction by NGS Software Requests},''
  \url{<https://www.ngs.noaa.gov/cgi-bin/grav_pdx.prl>}.

\bibitem{software} Software, installation instructions, CAD file, and an alternative circuit diagram to accompany this paper will be available shortly . 

\bibitem{Note3}
We deliberately use V\,s/m as units for
  $\protect \ensuremath {(BL)_\protect \mathrm {V}}$, to emphasize that V is a
  conventional unit, while m and s are SI units. With the unit of T\,m we would have lost this distinction.
  Analogously, we use the unit N/A for the quantity $(BL)_\mathrm{F}$.

\bibitem{Swanson10}
H.~E. Swanson and S.~Schlamminger, ``{Removal of zero-point drift from AB data
  and the statistical cost},'' {Meas. Sci.
  Technol.}
  \textbf{21}, 115104--115110 (2010).

\bibitem{Steiner06}
R.~L. Steiner, E.~R. Williams, D.~B. Newell, and R.~Liu, ``{Towards an
  electronic kilogram: An improved measurement of the Planck constant and
  electron mass},'' {Metrologia} \textbf{42}, 431--441 (2005).

\bibitem{Robinson12}
I.~Robinson, ``{Alignment of the NPL Mark II watt balance},'' {Meas. Sci.
  Technol.} \textbf{23}, 124012--124029 (2012).

\bibitem{Kibble14}
B.~Kibble and I.~Robinson, ``{Principles of a new generation of simplified and
  accurate watt balances},'' {Metrologia} \textbf{51}, S132--S139 (2014).
  
\bibitem{Knotts14}
S.~Knotts, P.~Mohr, and W.~Phillips, ``{An introduction to the New SI},'' arXiv:1503.03496 [physics.ed-ph], (2015).
  
\bibitem{facebook} Facebook page, LEGO Watt Balance
\url{<https://www.facebook.com/LEGOwattbalance>}




\end{thebibliography}
\end{document}